\documentclass[%
 reprint,
superscriptaddress,
showpacs,preprintnumbers,
 amsmath,amssymb,
 aps,
prb,
floatfix,
]{revtex4-1}

\usepackage{color}
\usepackage{graphicx}% Include figure files
\usepackage{dcolumn}% Align table columns on decimal point
\usepackage{bm}% bold math
\usepackage{hyperref}
\hypersetup{
    unicode=false,          % non-Latin characters in Acrobat’s bookmarks
    pdftoolbar=true,        % show Acrobat’s toolbar?
    pdfmenubar=true,        % show Acrobat’s menu?
    pdffitwindow=false,     % window fit to page when opened
    pdfstartview={FitH},    % fits the width of the page to the window
    pdftitle={My title},    % title
    pdfauthor={Author},     % author
    pdfsubject={Subject},   % subject of the document
    pdfcreator={Creator},   % creator of the document
    pdfproducer={Producer}, % producer of the document
    pdfkeywords={keyword1} {key2} {key3}, % list of keywords
    pdfnewwindow=true,      % links in new window
    colorlinks=true,       % false: boxed links; true: colored links
    linkcolor=blue,          % color of internal links (change box color with linkbordercolor)
    citecolor=blue,        % color of links to bibliography
    filecolor=blue,      % color of file links
    urlcolor=blue,       % color of external links
    plainpages=false        % hat Einfluss auf die Seitennummerierung
}

\begin{document}
\title{Nematic fluctuations and the magneto-structural phase transition in ${\rm Ba(Fe_{1-x}Co_x)_2As_2}$}
\date{\today}
\author{F. Kretzschmar}
\affiliation{Walther Meissner Institut, Bayerische Akademie der Wissenschaften, 85748 Garching, Germany}
\affiliation{Fakult\"at f\"ur Physik E23, Technische Universit\"at M\"unchen, 85748 Garching, Germany}
\author{T. B\"ohm}
\affiliation{Walther Meissner Institut, Bayerische Akademie der Wissenschaften, 85748 Garching, Germany}
\affiliation{Fakult\"at f\"ur Physik E23, Technische Universit\"at M\"unchen, 85748 Garching, Germany}
\author{U. Karahasanovi\'c}
\affiliation{Institute for Theoretical Condensed Matter Physics (TKM), Karlsruhe Institute of Technology (KIT), 76128 Karlsruhe, Germany}
\author{B. Muschler}
\altaffiliation{Present address: Zoller \& Fr\"ohlich GmbH, Simoniusstrasse 22, 88239 Wangen im Allg\"au, Germany}
\affiliation{Walther Meissner Institut, Bayerische Akademie der Wissenschaften, 85748 Garching, Germany}
\affiliation{Fakult\"at f\"ur Physik E23, Technische Universit\"at M\"unchen, 85748 Garching, Germany}
\author{A. Baum}
\affiliation{Walther Meissner Institut, Bayerische Akademie der Wissenschaften, 85748 Garching, Germany}
\affiliation{Fakult\"at f\"ur Physik E23, Technische Universit\"at M\"unchen, 85748 Garching, Germany}
\author{D. Jost}
\affiliation{Walther Meissner Institut, Bayerische Akademie der Wissenschaften, 85748 Garching, Germany}
\affiliation{Fakult\"at f\"ur Physik E23, Technische Universit\"at M\"unchen, 85748 Garching, Germany}
\author{J.~Schmalian}
\affiliation{Institute for Theoretical Condensed Matter Physics (TKM), Karlsruhe Institute of Technology (KIT), 76128 Karlsruhe, Germany}
\author{S.~Caprara}
\affiliation{Department of Physics, University of Rome ``Sapienza'', 00185 Roma, Italy}
\author{M. Grilli}
\affiliation{Department of Physics, University of Rome ``Sapienza'', 00185 Roma, Italy}
\author{C. Di\,Castro}
\affiliation{Department of Physics, University of Rome ``Sapienza'', 00185 Roma, Italy}
\author{J.~G. Analytis}
\altaffiliation{Current address: Department of Physics, University of California, Berkeley, CA 94720, USA}
\affiliation{Stanford Institute for Materials and Energy Sciences,
SLAC National Accelerator Laboratory, 2575 Sand Hill Road, Menlo Park, CA 94025, USA}
\affiliation{Geballe Laboratory for Advanced Materials \& Dept. of Applied Physics,
Stanford University, CA 94305, USA}
\author{J.-H. Chu}
\affiliation{Stanford Institute for Materials and Energy Sciences,
SLAC National Accelerator Laboratory, 2575 Sand Hill Road, Menlo Park, CA 94025, USA}
\affiliation{Geballe Laboratory for Advanced Materials \& Dept. of Applied Physics,
Stanford University, CA 94305, USA}
\author{I.~R. Fisher}
\affiliation{Stanford Institute for Materials and Energy Sciences,
SLAC National Accelerator Laboratory, 2575 Sand Hill Road, Menlo Park, CA 94025, USA}
\affiliation{Geballe Laboratory for Advanced Materials \& Dept. of Applied Physics,
Stanford University, CA 94305, USA}
\author{R. Hackl}
\affiliation{Walther Meissner Institut, Bayerische Akademie der Wissenschaften, 85748 Garching, Germany}
%\email{hackl@wmi.badw.de}

%%%%%%%%%%%%%%%%%%%%%%%%%%%%%%%%%%%%%%%%%%%%%%%%%%%%%%%%%%%%%%%%%%%%%%%%%%%%%%%%%%%%%%%%%
\begin{abstract}
  An inelastic light (Raman) scattering study of nematicity and critical fluctuations in ${\rm Ba(Fe_{1-x}Co_x)_2As_2}$ ($0\le x \le 0.051$) is presented. It is shown that the response from fluctuations appears only in $B_{1g}$ (${x^2-y^2}$) symmetry. The scattering amplitude increases towards the structural transition at $T_s$ but vanishes only below the magnetic ordering transition at $T_{\rm SDW} < T_s$, suggesting a magnetic origin of the fluctuations. The theoretical analysis explains the selection rules and the temperature dependence of the fluctuation response. These results make magnetism the favorite candidate for driving the series of transitions.%Below $T_{\rm SDW}$ the gap of the magnetically ordered phase opens up.
\end{abstract}
%%%%%%%%%%%%%%%%%%%%%%%%%%%%%%%%%%%%%%%%%%%%%%%%%%%%%%%%%%%%%%%%%%%%%%%%%%%%%%%%%%%%%%%%%
\maketitle
%%%%%%%%%%%%%%%%%%%%%%%%%%%%%%%%%%%%%%%%%%%%%%%%%%%%%%%%%%%%%%%%%%%%%%%%%%%%%%%%%%%%%%%%%

%\tableofcontents

\section{Introduction}
Nematic fluctuations and order play a prominent role in material classes such as the cuprates \cite{Ando:2002}, some ruthenates \cite{Borzi:2007} or the iron-based compounds \cite{Chu:2010,Chu:2012,Fernandes:2014,Kuo:2015} and may be interrelated with superconductivity \cite{Lederer:2015,Baek:2014,Gallais:2015,Fradkin:2015,Capati:2015}. In iron-based compounds \cite{Kamihara:2008,Rotter:2008} signatures of nematicity have been observed in a variety of experiments, and the magneto-structural phase transition is among the most thoroughly studied phenomena. When Fe is substituted by Co in ${\rm BaFe_2As_2}$ the structural transformation at $T_s$ precedes the magnetic ordering at $T_{\rm SDW} < T_s$  [Ref.~\onlinecite{Chu:2009}]. The nematic phase between $T_s$ and $T_{\rm SDW}$ is characterized by broken $C_4$ symmetry but preserved $O(3)$ spin rotational symmetry (no magnetic order). Nematic fluctuations are present even above $T_s$ in the tetragonal phase as has been demonstrated in studies of the elastic constants \cite{Bohmer:2014}. In strained samples, one observes orbital ordering in the photoemission spectra \cite{Yi:2011} and electronic nematicity by transport \cite{Chu:2012,Mirri:2014}. However, the fundamental question as to the relevance of the related spin \cite{Fernandes:2012a}, charge \cite{Gallais:2013} or orbital \cite{Kontani:2011,Kontani:2014,Baek:2014} fluctuations remains open. In fact, it is rather difficult to derive the dynamics and momentum dependence of the critical fluctuations with finite characteristic wavelengths \cite{Caprara:2005,Caprara:2015,Karahasanovic:2015} and to identify which of the ordering phenomena drives the instabilities.

Raman scattering provides experimental access to all types of dynamic nematicity but only the charge sector has been studied in more detail \cite{LeeWC:2009,Choi:2010,Gallais:2013}. However, also in the case of spin-driven nematic order the technique can play a prominent role for coupling to a two-spin operator whereas a four-spin correlation function is the lowest order contribution to the neutron cross section \cite{Fernandes:2014}. We exploit this advantage here and study the low-energy Raman response of ${\rm Ba(Fe_{1-x}Co_x)_2As_2}$ experimentally and interpret the results in terms of a microscopic model for a spin-driven nematic phase. % which includes the temperature dependence, the spectral shape, and the selection rules.
In addition to the  temperature dependence \cite{Gallais:2013} we address the spectral shape and the selection rules enabling us to explain the structural and magnetic transitions in a unified microscopic picture.

We study ${\rm Ba(Fe_{1-x}Co_x)_2As_2}$ single crystals having $x=0$, $x=0.025$, and $x=0.051$ as a function of photon polarization in the temperature range $4.2 < T \le 300$\,K. For the symmetry assignment we use the 1\,Fe unit cell making the fluctuations to appear in $B_{1g}$ symmetry. We use the appearance of twin boundaries and of the As $A_{1g}$ (${x^2+y^2}$) phonon line as internal thermometers for the structural and the magnetic phase transitions, respectively. In this way, $T_s$ and $T_{\rm SDW}$ can be determined with a precision of typically $\pm0.2$ and $\pm1$\,K, respectively.

\begin{figure*}[tbp]
  \centering
  \includegraphics[width=14cm]{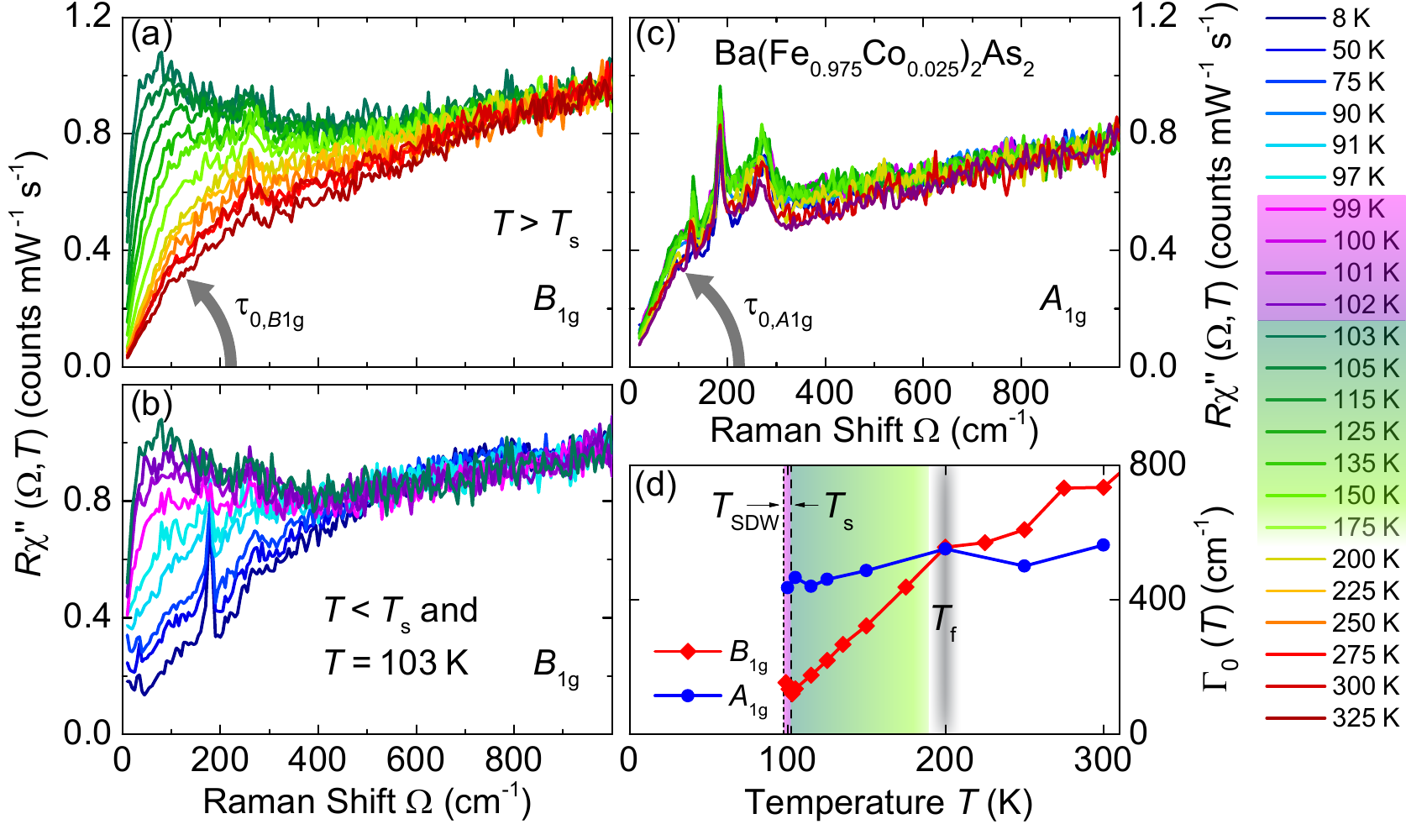}
  \caption{Raman response $R\chi^{\prime\prime}(\Omega,T)$ (raw data after division by the Bose-Einstein factor) of ${\rm Ba(Fe_{0.975}Co_{0.025})_2As_2}$ in (a) $B_{1g}$ above and (b) below $T_s$ and (c) $A_{1g}$ symmetry at temperatures as indicated. The initial slopes shown in (a) and (c) as grey arrows are proportional to the static two-particle lifetime in symmetry $\mu=A_{1g}$, ~$B_{1g}$. (d) Raman relaxation rates  $\Gamma_{0,\mu}(T)$ in $A_{1g}$ (blue circles) and $B_{1g}$ (red diamonds) symmetry as a function of temperature. The fluctuation range $T_s<T<T_f$  and the nematic phase $T_{\rm SDW}<T<T_s$ are indicated in green and magenta, respectively.
  }
  \label{fig:2.5Ts}
\end{figure*}

\section{Experiment}
The single crystals of undoped and Co-substituted ${\rm Ba(Fe_{1-x}Co_{x})_2As_2}$ were grown using a self-flux technique and have been characterized elsewhere \cite{Chu:2009}. The cobalt concentration was determined by microprobe analysis. $T_s$ and $T_{\rm SDW}$ are close to 134\,K in the undoped sample and cannot be distinguished. At nominally $x=0.025$ we find $T_s=102.8\pm0.1$\,K and $T_{\rm SDW}=98\pm1$\,K by directly observing the appearance of twin boundaries and a symmetry-forbidden phonon line, respectively (see Appendix~\ref{Asec:Ts} for details). The extremely sharp transition at $T_s$ having $\Delta T_s\approx 0.2$\,K indicates that the sample is very homogeneous in the area of the laser spot.

The experiments were performed with standard light scattering equipment. For excitation either a solid state laser (Coherent, Sapphire SF 532-155 CW) or an Ar ion laser (Coherent, Innova 300) was used emitting at 532 or 514.5\,nm, respectively. The samples were mounted on the cold finger of a He-flow cryostat in a cryogenically pumped vacuum. The laser-induced heating was determined experimentally (see Appendix~\ref{Asec:Ts}) to be close to 1\,K per mW absorbed power. The spectra represent the response $R\chi_\mu^{\prime\prime}(\Omega,T)$ ($\mu = A_{1g}$, $B_{1g}$, $A_{2g}$ and $B_{2g}$) that is obtained by dividing the measured (symmetry resolved) spectra by the Bose-Einstein factor $\{1+n(T,\Omega)\}=[1-\exp(-\hbar\Omega/k_BT)]^{-1}$. $\chi_\mu^{\prime\prime}(\Omega,T)$ is the imaginary part of the response function, and $R$ is an experimental constant that connects the observed photon count rates with the cross-section and the van Hove function and accounts for units. For simplicity the symmetry index $\mu$ is dropped in most of the cases. The symmetry selection rules refer to the 1\,Fe unit cell (see insert of Fig.~\ref{fig:T-spot}\,(b) in Appendix~\ref{Asec:Ts}) which is more appropriate for electronic and spin excitations.

\section{Experimental Results}
Fig.~\ref{fig:2.5Ts} shows the Raman response $R\chi^{\prime\prime}(\Omega,T)$ for ${\rm Ba(Fe_{0.975}Co_{0.025})_2As_2}$ for various temperatures in $A_{1g}$ and $B_{1g}$ (1\,Fe per unit cell) symmetry. $B_{2g}$ spectra were measured only at a few temperatures and found to be nearly temperature independent in agreement with previous data \cite{Gallais:2013}. Results for other doping levels $x$ are shown in Appendix~\ref{Asec:other}. The spectra comprise a superposition of  several types of excitations including narrow phonon lines and slowly varying continua arising from electron-hole (e-h) pairs; hence the continuum reflects the dynamical two-particle behavior. The $A_{1g}$ and $B_{1g}$ spectra predominantly weigh out contributions from the central hole bands and the electron bands, respectively \cite{Muschler:2009,Mazin:2010a}. The symmetry-dependent initial slope $\tau_{0,\mu}(T)$ ($\mu=A_{1g}$, $B_{1g}$, $B_{2g}$) [see Fig.~\ref{fig:2.5Ts}\,(a) and (c)] can be compared to transport data. $[\tau_{0,\mu}(T)]^{-1}$ corresponds to the static transport relaxation rate $\Gamma_{0,\mu}(T)$ of the conduction electrons \cite{Zawadowski:1990,Opel:2000,Devereaux:2007}. The memory function method facilitates the quantitative determination of the dynamic relaxation $\Gamma(\Omega, T)$ in absolute energy units \cite{Opel:2000}. The static limit can be obtained by extrapolation, $\Gamma_{0,\mu}(T) = \Gamma_{\mu}(\Omega \rightarrow 0, T)$ [see Appendix \ref{Asec:memo}]. In Fig.~\ref{fig:2.5Ts}\,(d) we show the result for $x=0.025$ corresponding to the spectra of Fig.~\ref{fig:2.5Ts}\,(a), (b) and (c). The results for all doping levels studied are compiled in Fig.~\ref{fig:ini_slope_all} in Appendix~\ref{Asec:memo} and compared to the scattering rates derived from the resistivities \cite{Chu:2009}.

Fig.~\ref{fig:2.5Ts}\,(d) displays one of the central results: Above approximately 200\,K $\Gamma_{0,\mu}(T)$ varies slowly and similarly in both symmetries. The more rapid decrease of $\Gamma_{0,B1g}(T)$ below 200\,K  is accompanied by a strong intensity gain in the range 20--200\,cm$^{-1}$ [see Fig.~\ref{fig:2.5Ts}\,(a)] as observed before in similar samples \cite{Choi:2010,Gallais:2013}. The intensity gain indicates that there is an additional contribution superposed on the e-h continuum which, as will be shown below, arises from fluctuations. Therefore, the kink in $\Gamma_{0,B1g}(T)$ is labeled $T_f$ and marks the crossover temperature below which nematic fluctuations can be observed by Raman scattering. At least for low doping, $T_f$ is relatively well defined. The kink allows us to separate the two regimes of the low-energy response above and below $T_f$ as being dominated by carrier excitations and fluctuations, respectively.

\begin{figure}[tbp]
  \centering
  \includegraphics[width=1.0\columnwidth]{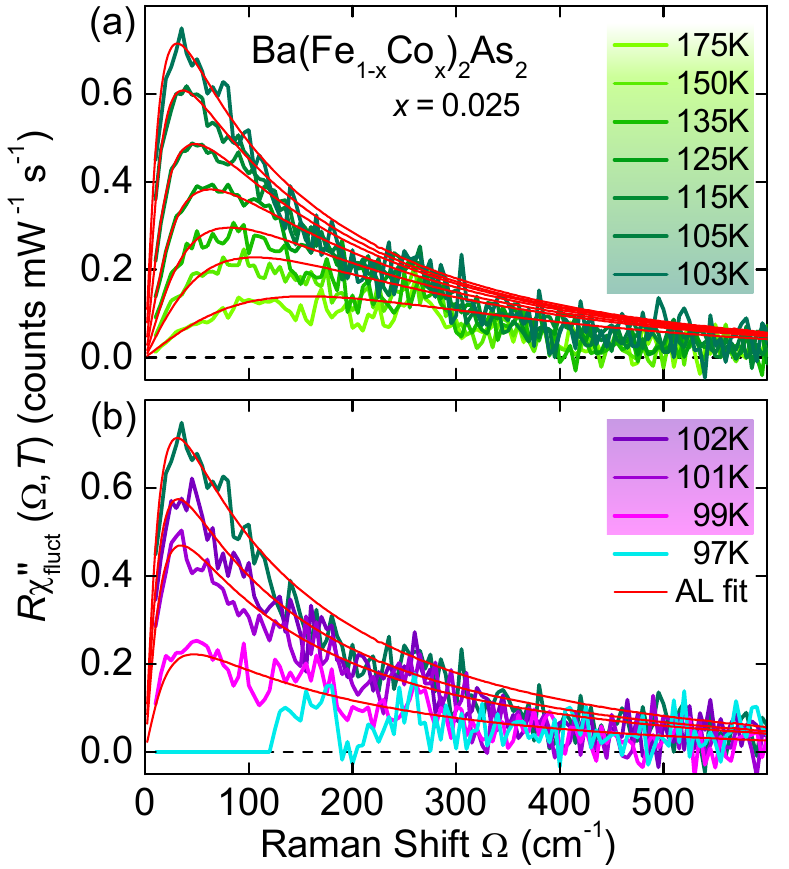}
  \caption{Fluctuation contribution to the Raman response of ${\rm Ba(Fe_{0.975}Co_{0.025})_2As_2}$ (a) above $T_s$ and (b) between $T_s$ and $T_{\rm SDW}$. The red lines are theoretical predictions on the basis of Aslamazov-Larkin diagrams\cite{Caprara:2005} describing the exchange of a pair of fluctuations (for details see Appendix~\ref{Asec:SR}).
  }
  \label{fig:2.5AL}
\end{figure}

The additional $B_{1g}$ signal below $T_f$  has to be treated in a way different from that in $A_{1g}$ symmetry and in $B_{1g}$ above $T_f$. Since it is rather strong it can be separated out with little uncertainty by subtracting the e-h continuum. We approximate the continuum at $T_f$ by an analytic function which is then determined for each temperature according to the variation of the resistivity and the $A_{1g}$ spectra and subtracted from all spectra at lower temperatures. The details are explained in Appendix~\ref{Asec:continuum}. The results of the subtraction procedure are shown in Fig.~\ref{fig:2.5AL}. The response increases rapidly towards $T_s$ without however diverging, and the maximum moves to lower energies.

As a surprise, the fluctuations do not disappear directly below $T_s$ [Fig.~\ref{fig:2.5AL}\,(b)] as one would expect if long-ranged order would be established. Rather, the intensity decreases continuously and the maximum stays approximately pinned implying that the correlation length does not change substantially between the two transitions at $T_s=102.8\pm0.2$\,K and $T_{\rm SDW}=98\pm1$\,K. The persistence of the fluctuations down to $T_{\rm SDW}$ strongly favors their magnetic origin.

\section{Theory}

\begin{figure}[h]
  \includegraphics[width=1\columnwidth]{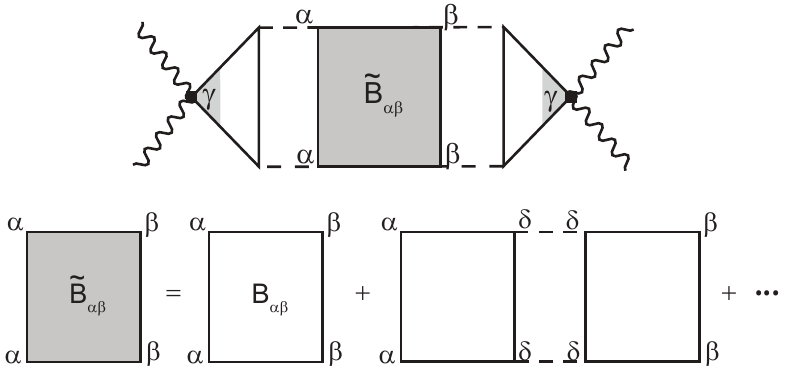}
  \caption{Higher order Aslamazov-Larkin diagrams for interacting fluctuations. The paramagnetic coupling between fluctuations mediated by fermions is obtained by inserting quaternions into the leading order diagram. The re-summed box $\tilde B_{\alpha \beta}$ is shaded grey. The first index of the matrix $B$ denotes the type $\alpha=X/Y$ of entering spin fluctuations, and the second index the type of exiting spin fluctuations.}
\label{fig:resum}
\end{figure}

We first compare the data to the theoretical model for thermally driven spin fluctuations associated with the striped magnetic phase ordering along ${\bf Q}_x=(\pi,0)$ or ${\bf Q}_y=(0,\pi)$. In leading order two noninteracting fluctuations carrying momenta $\bf Q$ and $-\bf Q$ are exchanged. Electronic loops (see Figs.~\ref{fig:resum} and \ref{fig:SR}) connect the photons and the fluctuations and entail $\bf Q$-dependent selection rules that were derived along with the response $R_{0,\mu}(\Omega)$ in Ref.~\onlinecite{Caprara:2005} and are summarized in Appendix~\ref{Asec:SR}. In brief, since the response results from a sum over all electronic momenta close to the Fermi surface cancellation effects may occur if $\bf Q$ connects parts on different Fermi surface sheets having form factors $\gamma_\mu(\bf k)$ with opposite sign.  For the ordering vectors $(\pi,0)$ and $(0,\pi)$ the resulting selection rules explain the enhancement of the signal in $B_{1g}$ symmetry and its absence in the $A_{1g}$ and $B_{2g}$ channels. In contrast, for ferro-orbital ordering with ${\bf Q}=(0,0)$ as found in FeSe \cite{Baek:2014} the fluctuation response would appear in all symmetries.

However, the lowest-order diagrams alone can only account for the spectral shape whereas the variation of the intensity around $T_s$ remains unexplained. In order to describe this aspect, we consider the interaction of fluctuations among themselves and with the lattice, all of which becomes crucial in the treatment of spin-driven nematicity \cite{Karahasanovic:2015}.

The interactions between spin fluctuations can be represented by a series of quaternion paramagnetic couplings mediated by fermions inserted into the leading order Aslamazov-Larkin diagrams as shown in Fig.~\ref{fig:resum}.  The inserted fermionic boxes effectively resemble the dynamic nematic coupling constant $g$ of the theory.

We have analyzed the problem by extending $SU(2) \rightarrow SU(N)$ and taking the large $N$ limit. For small frequencies $\Omega$ and in the large-$N$ limit, after re-summing an infinite number of such box-like Aslamazov-Larkin diagrams, the Raman response function $\tilde R_{B1g}(\Omega)$ reads,
\begin{eqnarray}
  \label{eq:raman}
  \tilde R_{B1g}(\Omega)&=&R_{0,{B1g}}(\Omega) \left [ 1+g \chi^{\rm el}_{\rm nem} (0) \right ].
\end{eqnarray}
Eq.~\eqref{eq:raman} states that the Raman response is proportional to the electronic contribution to the susceptibility of the nematic order parameter,
\begin{eqnarray}
  \label{eq:chinemel}
  \chi^{\rm el}_{\rm nem}(0)&=&\frac{\int_{q}  \chi_{\rm mag}^2(q)}{1-g\int_{q} \chi_{\rm mag}^2(q)}.
\end{eqnarray}
$\chi_{\rm mag}(q)$ represents the magnetic susceptibility that diverges at $T_{\rm SDW}$. For $g\neq0$ $\chi^{\rm el}_{\rm nem}(0)$ has a Curie-like $|T-T^\ast|^{-1}$ divergence at $T^\ast\ge T_{\rm SDW}$.

If the spins (or charges) couple to the lattice the susceptibility of the nematic order parameter is given by \cite{Chu:2012,Kontani:2014,Karahasanovic:2015}
\begin{eqnarray}
\chi_{\rm nem}(0)=\frac{\int_{q}  \chi_{\rm mag}^2(q)}{1-\left [ g +({\lambda_{\rm sl}^2}/{c_0^{\rm s}})\right ]\int_{q} \chi_{\rm mag}^2(q)},
\label{eq:chinem}
\end{eqnarray}
where $\lambda_{\rm sl}$ denotes the magneto-elastic coupling, and $c_0^{\rm s}$ is the bare elastic constant. Obviously, $\chi_{\rm nem}(0)$ diverges at higher temperature than $\chi^{\rm el}_{\rm nem}(0)$. We identify $T_s \ge T^\ast$ with the structural transition and conclude that the Raman response (Eq.~\ref{eq:raman}) develops only a maximum rather than a divergence at $T_s$ in agreement with the experiment.

Close to $T_s$, we expect Eq.~\eqref{eq:raman} to hold qualitatively also inside the nematic phase, $T_{\rm SDW}<T<T_s$. We argue \cite{Fernandes:2012a} that $\chi^{\rm el}_{\rm nem} (0)$ and, according to Eq.~\eqref{eq:raman}, the Raman amplitude is smaller than in the disordered (tetragonal) state but different from zero. This explains the continuous reduction of the Raman response of spin fluctuations upon entering the nematic state. One can also show that the $A_{1g}$ response gets even further suppressed if one includes collisions between the fluctuations.

\section{Discussion}

\begin{figure}[tbp]
  \centering
  \includegraphics[width=1.0\columnwidth]{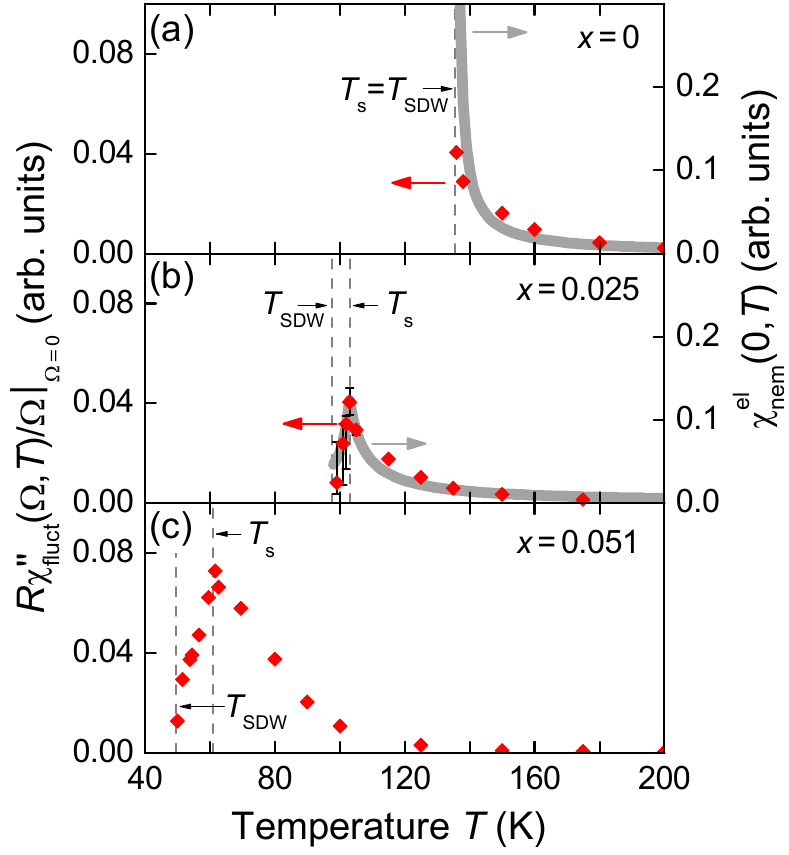}
  \caption{Temperature dependence of the initial slope of the fluctuation response. The initial slope is given in arbitrary units since only the temperature dependence matters. The error bars represent the differences originating in the subtraction of the e-h continuum. The grey curves represent $|T-T^\ast|^{-1}$ with $T^\ast<T_s$ (see text).
  }
  \label{fig:ini_slope}
\end{figure}

As shown in Eq.~\eqref{eq:raman} the full Raman response $\tilde R_\mu(\Omega)$ is proportional to the bare response $R_{0,\mu}(\Omega)$ and to the electronic nematic susceptibility $\chi^{\rm el}_{\rm nem}(0)$. Hence, the spectral shape is essentially given by $R_{0,\mu}(\Omega)$, that is therefore used in Fig.~\ref{fig:2.5AL} to fit the data, whereas the intensity is dominated by the prefactor $|T-T^\ast|^{-1}$. Since the theoretical model is valid only in the limit of small frequencies we argue that the initial slope reflects the temperature dependence of the intensity and is proportional to $\chi^{\rm el}_{\rm nem}(0)$, at least close to the transition. For generally reflecting the spectral shape above $T_{\rm SDW}$ (Eq.~\ref{eq:raman}), $R_{0,B1g}(\Omega,T)$ enables us to directly extract the initial slope of the experimental spectra by plotting $R_{0,B1g}(\Omega,T)/\Omega$ for all temperatures (see Appendix~\ref{Asec:slope}). These results are compiled in Fig.~\ref{fig:ini_slope} along with the variation of $\chi^{\rm el}_{\rm nem}(0,T)$ expected from mean-field theory. For low doping, we find qualitative agreement in the ranges $T_{\rm SDW}<T<T_s$ and $T_s<T$. For higher doping the interactions between fluctuations become dominant and the mean field prediction breaks down [Fig.~\ref{fig:ini_slope}\,(c)].

\begin{figure}[tbp]
  \centering
  \includegraphics[width=1.0\columnwidth]{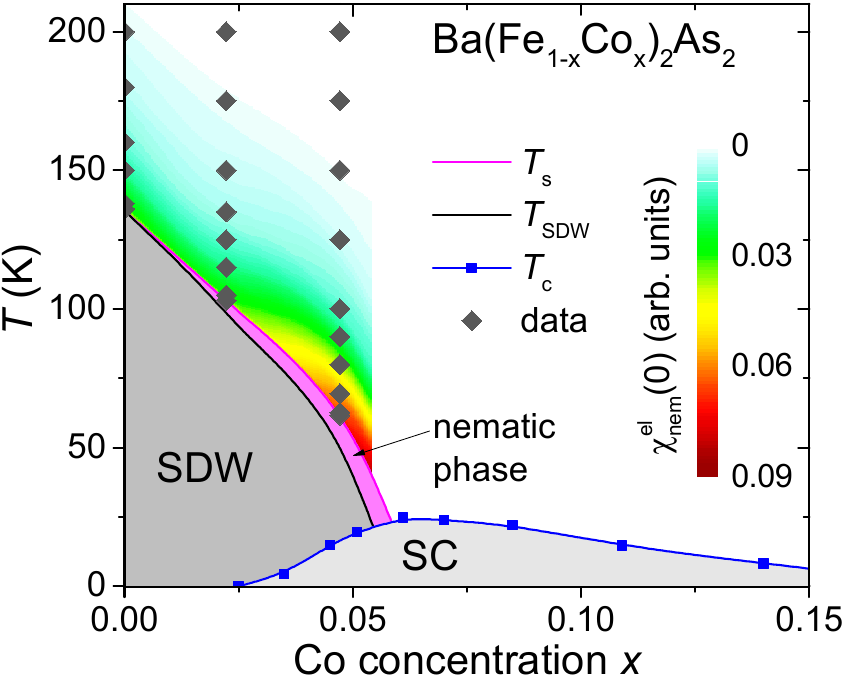}
  \caption{Phase diagram of ${\rm Ba(Fe_{1-x}Co_{x})_2As_2}$. The full lines limiting the nematic phase (magenta) and the blue squares representing the transition temperature $T_c$ of superconducting samples were derived in Ref.~\onlinecite{Chu:2009}. Grey diamonds represent doping and temperature positions of the current Raman data. The red-green field between $T_{s}$ and $T_f$ represents the initial slope of the spectra according to the color scale on the right.
  }
  \label{fig:pd}
\end{figure}

The fluctuations were also studied at various other doping levels in the range $0\le x \le 0.085$. Up to 6.1\% Co substitution fluctuations were observed. In contrast to other publications \cite{Gallais:2013} we were not able to clearly identify and isolate the response from fluctuations at 8.5\%. The results up to 5.1\% are unambiguous and are compiled in
Fig.~\ref{fig:pd}.  The fluctuations can be observed over a temperature range of  approximately 70--100\,K. This is more than in most of the other experiments on unstrained samples and comparable to what is found in the cuprates \cite{Muschler:2010a,Caprara:2015}.

\section{Conclusions}
The detailed experimental and theoretical study of  the light scattering response in ${\rm Ba(Fe_{1-x}Co_{x})_2As_2}$ reveals a broad range of spin fluctuations obeying $B_{1g}$ selection rules. The selection rules can be explained only for a finite ordering vector ${\bf Q} = (\pi,0)$. Any type of order with ${\bf Q} = (\pi,\pi)$ or ${\bf Q} = (0,0)$ such as ferro-orbital order is not compatible with the experiment. By observing the twin boundaries and the As phonon intensity we are able to determine the structural and magnetic transition temperatures with unprecedented precision. This observation allows us to conclude that the intensity of the fluctuation response is maximal at $T_s$ and vanishes at $T_{\rm SDW}$. The divergence of the intensity expected at $T_s$ from the electronic nematic susceptibility alone is shifted to lower temperature due to magneto-elastic coupling. Therefore, only an intensity maximum is observed at $T_s$. This fact together with the observation that the signal disappears at $T_{\rm SDW}$ supports the spin-driven nematic phase scenario. Hence magnetism is likely to be behind the transitions at least in ${\rm Ba(Fe_{1-x}Co_{x})_2As_2}$ and makes its fluctuations a candidate for driving superconductivity.

\begin{acknowledgments}
We acknowledge useful discussions with T.\,P. Devereaux, Y. Gallais, S.\,A. Kivelson, B. Moritz, and I. Paul. Financial support for the work came from the DFG via the Priority Program SPP\,1458 (project nos. HA\,2071/7 and SCHM\,1035/5), from the Bavarian Californian Technology Center BaCaTeC (project no. A5\,[2012-2]), and from the Transregional Collaborative Research Center TRR\,80. U.K. and J.S. were supported by the Helmholtz Association, through the Helmholtz post-doctoral grant PD-075 'Unconventional order and superconductivity in pnictides'.
\end{acknowledgments}

\begin{appendix}
\label{sec:appendix}

\section{Determination of the spot temperature}
\label{Asec:Ts}
\begin{figure}[tbp]
  \centering
  \includegraphics[width=1.0\columnwidth]{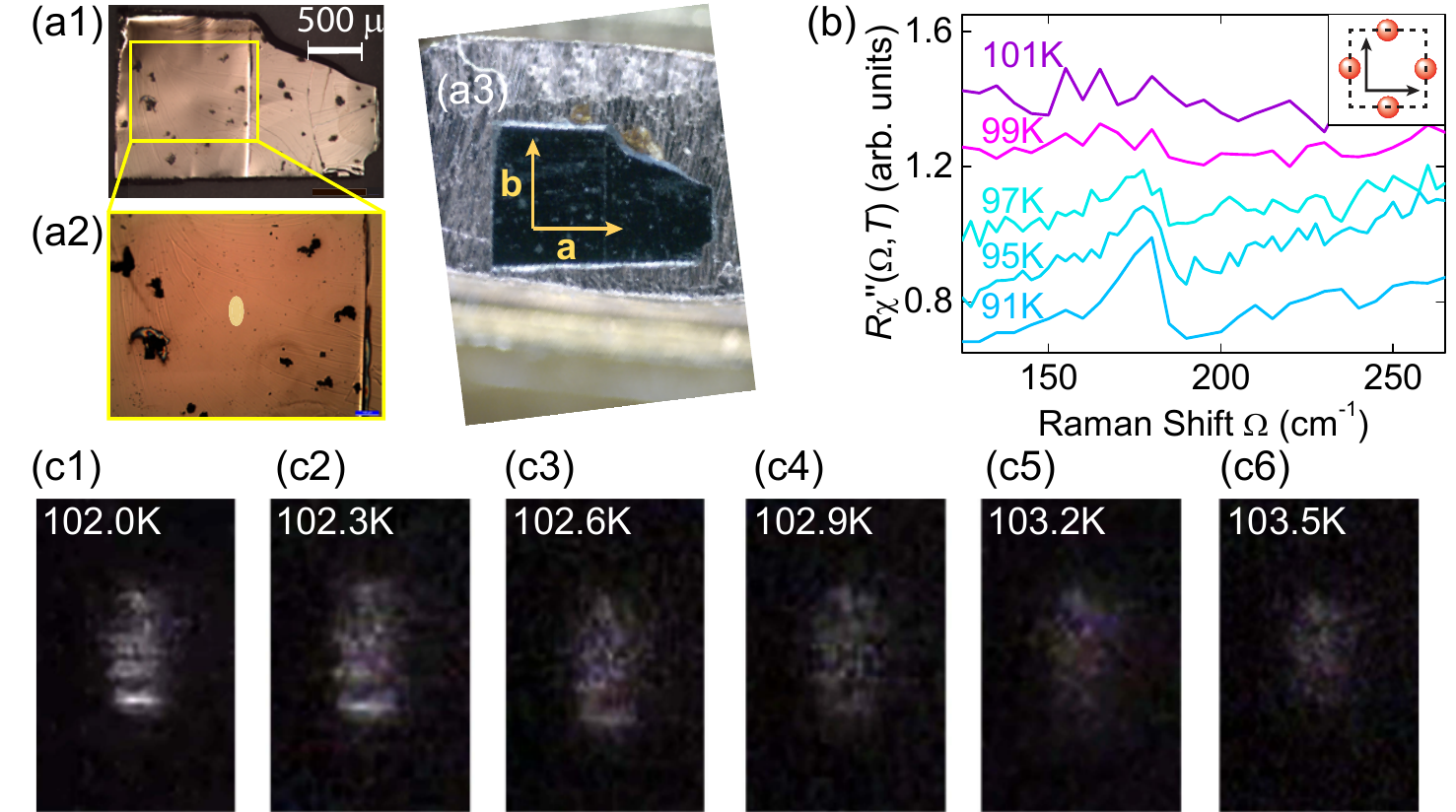}
  \caption{Determination of the spot temperature. (a1) and (a2) show the sample surface and a zoom in thereof. In (a2) the approximate laser spot is indicated schematically. (a3) shows the crystallographic axes of the tetragonal phase. (b) $A_g$ phonon around $T_{\rm SDW}$. Above $T_{\rm SDW}$ the symmetry leakage of the phonon is negligibly small. Only below $T_{\rm SDW}$ the intensity becomes appreciable. (c) Image of the illuminated spot as a function of temperature. The horizontal lines in (c1)--(c3) result from twin boundaries. In addition to the appearance of twin boundaries the reversible adsorption of residual gas atoms and molecules starts instantaneously and enhances the stray-light in the spot.
  }
  \label{fig:T-spot}
\end{figure}

\begin{figure}[tbp]
  \centering
  \vspace*{2mm}
  \includegraphics[width=1.0\columnwidth]{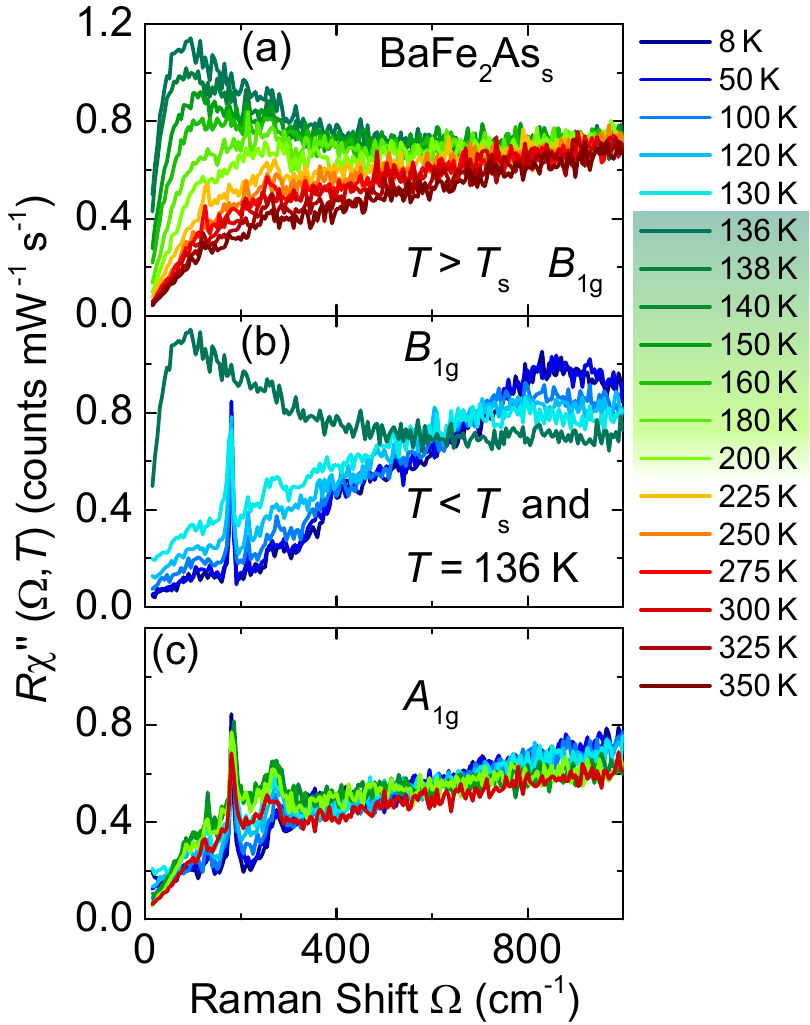}
  \caption{Raman response $R\chi^{\prime\prime}(\Omega,T)$ (raw data) of ${\rm BaFe_2As_2}$ in (a), (b) $B_{1g}$ and (c) $A_{1g}$ symmetry  above and below the structural transition $T_s$ at temperatures as indicated.
  }
  \label{fig:0Ts}
\end{figure}

\begin{figure}[tbp]
  \centering
  \vspace*{2mm}
  \includegraphics[width=1.0\columnwidth]{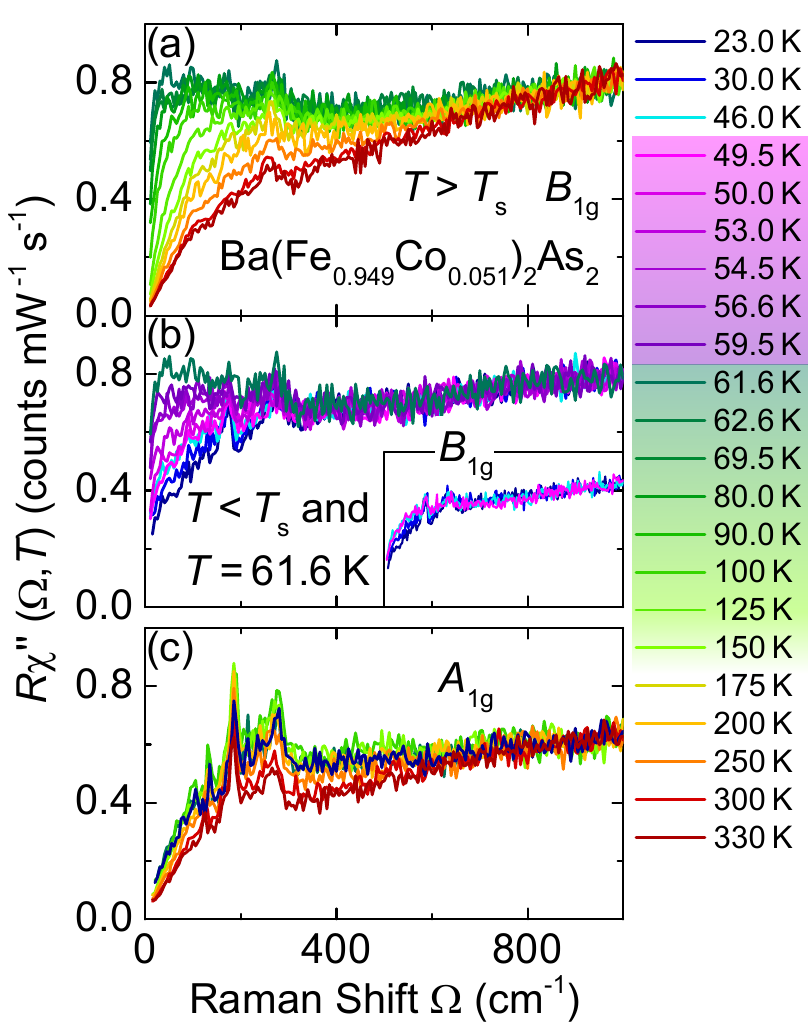}
  \caption{Raman response $R\chi^{\prime\prime}(\Omega,T)$ (raw data) of ${\rm Ba(Fe_{0.949}Co_{0.051})_2As_2}$ in (a), (b) $B_{1g}$ and (c) $A_{1g}$ symmetry  above and below the structural transition $T_s$ at temperatures as indicated. The inset in (b) shows that the SDW gap starts opening within 5\,K below $T_{\rm SDW}$.
  }
  \label{fig:5.1Ts}
\end{figure}

In Figs.~\ref{fig:2.5AL}, \ref{fig:0Ts} and \ref{fig:5.1Ts} we show that the response from fluctuations is maximal at $T_s$ and then decreases. For $x=0$ the decrease is very rapid, at $x=0.025$ and 0.051 the fluctuations disappear only below $T_{\rm SDW}$. Since $\Delta T=T_s-T_{\rm SDW}$ is small close to zero doping, the laser-induced heating has to be determined precisely. In addition, a large temperature gradient in the spot would lead to a substantial reduction of the maximal fluctuation intensity. Great care was therefore taken to keep the temperature gradient in the spot small and to determine the spot temperature and to calibrate it against intrinsic thermometers. The calibration is possible since twin boundaries develop below $T_s$ in the samples with $x=0.025$ and 0.051 facilitating a very precise determination of $T_s$. First we studied the effect of increasing laser power $P_L$ at different holder temperatures $T_h$ on the twin pattern that can be seen, e.g., in Fig.~\ref{fig:T-spot}\,(c1). In this way the laser heating $\Delta T_L$ was determined to be $1\pm0.1$\,K/mW for a spot diameter $d=50\,\mu$m. (Note that $\Delta T_L$ scales as $d^{-1}$ and not as $d^{-2}$.) Next we heat the sample slowly through $T_s$ using $P_L=0.3$\,mW as shown in a series of snapshots in Fig.~\ref{fig:T-spot}\,(c1)--(c6). The twin boundaries appear as horizontal lines and are most pronounced in (c1). With increasing temperature they ``melt'' and finally disappear completely at 102.9\,K (extrapolated sample temperature for $P_L=0$), and we identify $T_s=102.9$\,K. %The transition can be observed very clearly in movie M1.

For estimating $T_{\rm SDW}$ we analyze the phonons. The $A_{1g}$ As vibration was reported to appear in $B_{2g}$ symmetry below $T_s$ \cite{Chauviere:2009}. (We maintain the tetragonal 2\,Fe unit cell here as opposed to the main text to avoid confusion with the usual phonon assignment. In the proper orthorhombic 4\,Fe unit cell applying below $T_{\rm SDW}$ the phonon switches to $A_g$ symmetry, and $B_{2g}$ symmetry is not accessible any further with in-plan polarizations.) Our precise temperature determination shows for $x=0.025$ that the anomalous intensity does not appear at $T_s$. Rather the phonon anomaly appears only at approximately 97\,K as shown in Fig.~\ref{fig:T-spot}\,(b). According to the phase diagram the magnetic transition is offset by approximately 4-5\,K at $x=0.025$. This is actually not unexpected for a phonon that is not coupled to the lattice distortion by symmetry \cite{Miller:1967}. By measuring the $B_{2g}$ intensity of the $A_{1g}$ phonon we can therefore identify the magnetic transition temperature and find $T_{\rm SDW}=98\pm1$\,K.

For $x=0.051$ we find $T_s=61.0\pm0.2$\,K and $T_{\rm SDW}=51\pm2$\,K. Here, the $A_{1g}$ phonon appears already above $T_{\rm SDW}$, and we identify $T_{\rm SDW}$ with the strongest increase of the intensity. In addition, we know the width of the nematic phase from the phase diagram \cite{Chu:2009} (Fig.~\ref{fig:pd}) and find an anomaly of $\Gamma_{0,\mu}(T)$ close to  $T_{\rm SDW}$ [see Fig.~\ref{fig:ini_slope_all}\,(d)]. Hence the relevant temperatures are known with high precision.

\section{Results at $x=0$ and $x=0.051$}
\label{Asec:other}
\begin{figure}[tbp]
  \centering
  \includegraphics[width=0.95\columnwidth]{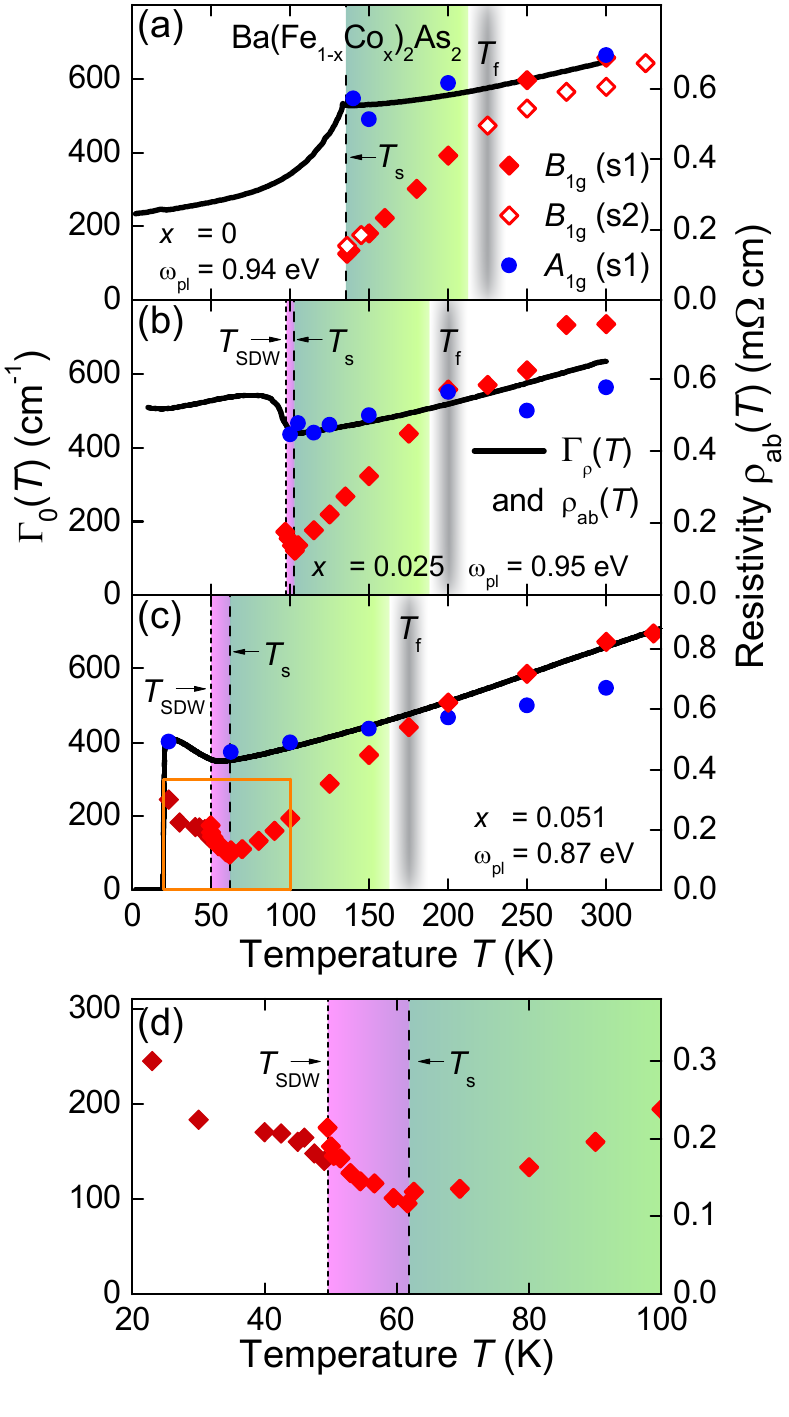}
  \caption{Static Raman relaxation rates $\Gamma_0(T)$ of (a) ${\rm BaFe_2As_2}$, (b) ${\rm Ba(Fe_{0.975}Co_{0.025})_2As_2}$, and (c) ${\rm Ba(Fe_{0.949}Co_{0.051})_2As_2}$. $\Gamma_0(T)$ is derived formally via the memory function method\cite{Opel:2000} as described in appendix~\ref{Asec:memo}. Above the onset temperature of the fluctuations $T_f$ the results in both $A_{1g}$ (blue) and $B_{1g}$ (red) symmetry return results similar  to those from the resistivity\cite{Chu:2009} (right ordinate). Using a Drude model, the resistivities $\rho(T)$ can be converted into scattering rates. At $T_f$ the temperature dependence in the $B_{1g}$ symmetry becomes much stronger. (d) If the resolution in temperatures is very high, one finds anomalies of $\Gamma_{0,\mu}(T)$ at $T_s$ and $T_{\rm SDW}$ which facilitates the independent determination of $T_s$ and $T_{\rm SDW}$ directly from the electronic Raman spectra.
  }
  \label{fig:ini_slope_all}
\end{figure}

Figs.~\ref{fig:0Ts} and \ref{fig:5.1Ts} show the experimental results for $x=0$ and $x=0.051$. At $x=0$ the two transitions $T_{\rm SDW}$ and $T_s$ either coincide or are too close to be observed separately while the response of the SDW phase can be identified clearly as observed earlier \cite{Chauviere:2010,Sugai:2012}. At $x=0.051$ the fluctuations can be separated out in the usual way as described below. If an extraction is attempted in a similar way at $x=0.085$ the variation with temperature cannot be described with Aslamazov-Larkin-type of fluctuations. Although the response increases slightly towards lower temperature \cite{Gallais:2013} and the elastic constants may still indicate an instability up to 9\% Co substitution \cite{Bohmer:2014} we do not feel comfortable to extract parameters in this case. The results for $\Gamma_{0,\mu}(T)$ are compiled in Fig.~\ref{fig:ini_slope_all}.

\begin{figure}[tbp]
  \centering
  \vspace{1mm}
  \includegraphics[width=0.8\columnwidth]{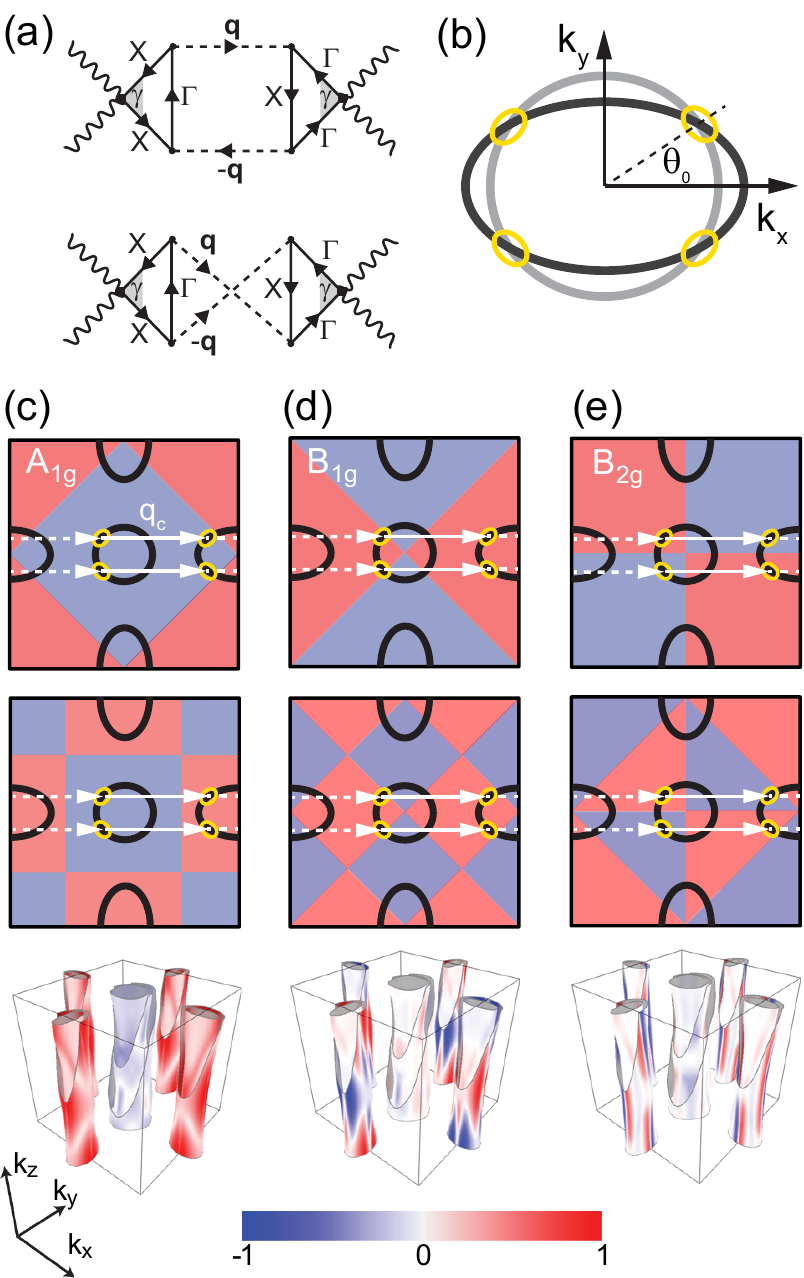}
  \caption{Scattering from fluctuations. (a) Examples of Aslamazov-Larkin diagrams describing light scattering from critical fluctuations  with momentum $\pm\bf q_c$ and energy $\omega_m$. The dashed lines represent the magnetic fluctuations, the full lines the fermionic propagators. (b) Hole- (grey) and back-folded electron-like (black) Fermi surfaces intersecting in the hot-spots (yellow circles). The selection rules can be deduced by considering cancelation effects arising from different hot-spot contributions inside the fermionic loops described in Eq.~\eqref{eq:f_loop}. The first and second row of (c), (d) and (e) show the signs and nodes of the first and second order $A_{1g}$, $B_{1g}$, and $B_{2g}$ Brillouin zone harmonics that indicate where cancellation effects can and cannot be expected. The $\bf q_c$ vectors for ($\pi,0$) and equivalent fluctuations are indicated by full and broken arrows, respectively. The last row shows the vertices derived from the second derivative of tight-binding band structure (effectice mass approximation) of Graser \textit{et al.} (Ref.~\onlinecite{Graser:2010b}). These vertices provide the best estimate for the sensitivity on the Fermi surface \cite{Devereaux:2007}. The $A_{1g}$ vertices for the hole and the electron bands are predominantly negative (blue) and positive (red), respectively.  The effective mass approximation shows that the $A_{1g}$ response will be dominated by the second order vertex $\cos k_x\cos k_y$ rather than the lowest order one as already pointed out in Ref.~\onlinecite{Mazin:2010a}.
  }
  \label{fig:SR}
\end{figure}

\section{Memory function and static relaxation rates}
\label{Asec:memo}
In Fig.~\ref{fig:2.5Ts}\,(d) symmetry-dependent static relaxation rates $\Gamma_{0,\mu}(T)$ are shown for $\mu=A_{1g}$ and $B_{1g}$,
\begin{equation}
  \frac{\hbar}{\tau_{0,\mu}(T)}=\Gamma_{0,\mu}(T)=\left.\left(\frac{\partial R\chi^{\prime\prime}_\mu(\Omega,T)}{\partial\Omega}\right)^{-1}\right|_{\Omega=0}.
\label{eq:slope}
\end{equation}
Since the overall intensity of the spectra is not known in absolute units the experimental constant $R$, to which the initial slope ${\tau_{0,\mu}(T)}$ is proportional, cannot be pinned down. Therefore, one needs additional information if one is interested in energy units for $\Gamma_{0,\mu}(T)$. Only then $\Gamma_{0,\mu}(T)$ can be compared to transport data. This problem was solved a while ago by adopting the memory function method \cite{Gotze:1972,Allen:1977} for Raman scattering \cite{Opel:2000}. Then $\Gamma_{0,\mu}(T)$ can be derived by extrapolating the dynamic Raman relaxation rates $\Gamma_\mu(\Omega,T)=\hbar/\tau_\mu(\Omega,T)$. The results for all doping levels are compiled in Fig.~\ref{fig:ini_slope_all}.

If a Drude model is applied  the resistivities $\rho(T)$ can be converted into static scattering rates. Using a plasma frequency close to 1\,eV in rough agreement with optical data\cite{Drechsler:2010}, the analysis shows that the Raman and transport results are compatible above a doping dependent temperature $T_f$ that is identified here with the onset of fluctuations in agreement with results from other methods. Transport and Raman scattering agree to within the experimental precision, possibly indicating the common origin of the electronic relaxation on the electron and hole bands.

\section{Aslamazov-Larkin Diagrams and Selection Rules}
\label{Asec:SR}
The coupling of visible light to critical fluctuations with wavevectors $|\bf q_c|={\bf Q}>0$ and energy (mass) $\omega_m$ is possible only via the creation of two excitations with opposite momenta warranting zero net momentum transfer applying for photon energies in the eV range [Fig.~\ref{fig:SR}\,(a)]. This process can be described by Aslamazov-Larkin (AL) diagrams \cite{Caprara:2005}. We assume a simplified model of the Fermi surface. The central sheet is a circular hole-like pocket around the $\Gamma$ point [grey circle in Fig.~\ref{fig:SR}\,(b)]. The two electron-like elliptical pockets with the  principle axes rotated by $90^{o}$ are centered at the $X$ $(\pm\pi,0)$ and $Y$ $(0,\pm\pi)$ points of the 1\,Fe BZ. If they are backfolded they intersect with the central hole band as indicated by yellow circle in [Fig.~\ref{fig:SR}\,(b)]. The fluctuation contribution to the Raman spectrum has been analyzed by Caprara and coworkers for the cuprates \cite{Caprara:2005} and arises from the AL diagrams shown in Fig~\ref{fig:SR}\,(a). The selection rules can be deduced by considering cancelation effects arising from different hot-spots within the fermionic loop as shown in Fig.~\ref{fig:SR}\,(a). Even if the entire Fermi surface is taken into account the selection rules still work in the Fe-based materials. For instance, in either case full cancellation is found for $B_{2g}$ symmetry\cite{Karahasanovic:2015}.

Explicitly written out, the fermionic loop is given by \cite{Venturini:2000,Caprara:2005,Caprara:2015,Karahasanovic:2015}
\begin{eqnarray}
  \label{eq:f_loop}
  \theta_{i,\mu}({\mathbf{q}_c},\Omega,\omega_m) &=& \theta_{i,\mu}^{(1)}({\mathbf{q}_c},\Omega,\omega_m)+\theta_{i,\mu}^{(2)}({\mathbf{q}},\Omega,\omega_m),
  \nonumber \\
  \theta_{i,\mu}^{(1)}({\mathbf{q}_c},\Omega,\omega_m) &=&T\sum \limits_{n} \int_{{\mathbf{k}}}
  \gamma^{\mu}_{{\mathbf{k}}}G_{\Gamma}({\mathbf{k}},\varepsilon_{n}- \Omega)G_{\Gamma}({\mathbf{k}},\varepsilon_{n}) \nonumber \\
  & &\times G_{i}({\mathbf{k}}-{\mathbf{q}_c}, \varepsilon_{n}- \omega_m), \nonumber \\
  \theta_{i,\mu}^{(2)}({\mathbf{q}_c},\Omega,\omega_m)&=&T\sum \limits_{n} \int_{{\mathbf{k}}}
  \gamma^{\mu}_{{\mathbf{k}}}G_{i}({\mathbf{k}},\varepsilon_{n}- \Omega)G_{i}({\mathbf{k}},\varepsilon_{n}) \nonumber \\
  & &\times G_{\Gamma}({\mathbf{k}}-{\mathbf{q}_c}, \varepsilon_{n}- \Omega + \omega_m),
\end{eqnarray}
where $\gamma^{\mu}_{\mathbf{k}}$ is the form factor ($\mu=B_{1g},~A_{1g}$ etc.), and $G_i$ is the electron propagator on band $i=\Gamma,~X,~Y$. $\varepsilon_{n}$ is the electronic energy and $\Omega$ is the energy difference between the incoming and scattered photons. Experimentally, pure symmetries can be obtained from linear combinations of the response measured at appropriate polarizations of the incoming and scattered photons $\hat e_i$ and $\hat e_s$.

\begin{figure*}[tbp]
  \centering
  \vspace*{2mm}
  \includegraphics[width=14cm]{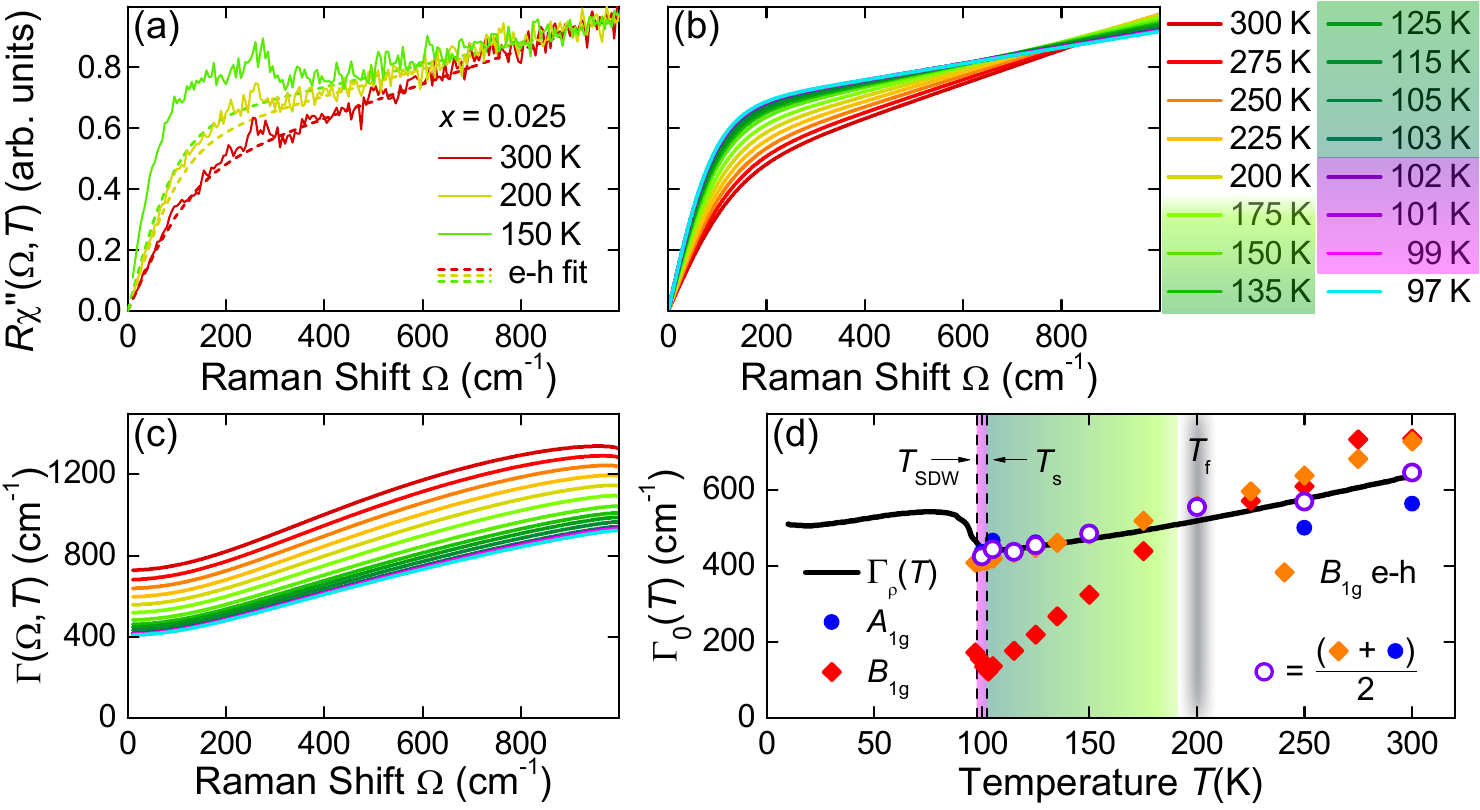}
  \caption{Determination of the e-h continuum and subtraction from the full response in ${\rm Ba(Fe_{1-x}Co_{x})_2As_2}$ ($x=0.025$). (a) The analytical function (red dashes) provides a reasonable fit at 300\,K. At lower temperatures the fluctuations emerge above the continuum and the analytical functions lie below the data (yellow and green dashes). (b) The analytical model is varied so as to reproduce the temperature dependence of the resistivity as shown as orange diamonds in panel (d). (c) Dynamical relaxation rates $\Gamma(\Omega,T)$ derived from the synthetic spectra in panel (b). The zero-energy extrapolation values of  $\Gamma_c(0,T)$ are plotted as orange diamonds in (d). The $A_{1g}$ and $B_{1g}$ data are taken from Fig.~\ref{fig:2.5Ts}.
  }
  \label{fig:synthetic_e-h}
\end{figure*}

For illustration purposes the fermionic loop $\theta$ is approximated in the hot-spot approximation. Hot-spots are regions in momentum space where both $\bf k$ and $\bf k\pm \bf q_c$ lie on the Fermi surface [Fig \ref{fig:SR}\,(b)]. Since the loop $\theta$ contains the symmetry factor $\gamma(\bf k)$ linearly inside the momentum integral the sign of $\gamma(\bf k)$ is crucial. If $\gamma(\bf k)$ changes sign for different hot spots connected by $\bf q_c$ (Fig.~\ref{fig:SR}\,(c), (d), and (e) for $A_{1g}, ~B_{1g}, ~B_{2g}$, respectively) there will be full or partial cancelation within $\theta$. Full cancelation is observed for the first two (and also higher) orders of $B_{2g}$ symmetry [Fig.~\ref{fig:SR}\,(e)]. In contrast, $\gamma(\bf k)$ does not change sign across different hot-spots for the $B_{1g}$ channel. Consequently, in $B_{1g}$ and $B_{2g}$ the fluctuations are Raman active and inactive, respectively.

The $A_{1g}$ symmetry is more complicated in that the first order contribution, proportional to $\cos(k_x)+\cos(k_y)$ [upper row of Fig.~\ref{fig:SR}\,(c)], would be as strong as the $B_{1g}$ contribution [Fig.~\ref{fig:SR}\,(d)] whereas the second order contribution ($\cos(k_x)\cos(k_y)$) [second row of Fig.~\ref{fig:SR}\,(c)] shows cancelation. For clarifying the relative magnitude of the two orders we analyze the effective mass vertices on the Fermi surfaces (second derivative or curvature of the band structure), that are the best approximations for the sensitivity away from resonances, in a way similar to what was proposed in Ref.~\onlinecite{Mazin:2010a}. The last row of Fig.~\ref{fig:SR}\,(c) shows that the band curvatures corresponding to the $A_{1g}$ vertex
\begin{equation}
  \gamma_{i,A1g}({\bf k})=
  \frac{\partial^2 \varepsilon_{i,\bf k}}{\partial k_x\partial k_x}+
  \frac{\partial^2 \varepsilon_{i,\bf k}}{\partial k_y\partial k_y}
\end{equation}
on the Fermi surface of the hole and the electron bands $(i)$ are predomininantly negative and positive, respectively, as expected already for simple parabolic bands with masses $m_h \approx -m_e$ although there are various near nodes on both bands. This result shows that $\cos(k_x)\cos(k_y)$ is the leading order. We note that $\cos(k_x)\cos(k_y)$ predicts a stronger mixing of the particle-hole response from the electron and hole bands than $\cos(k_x)+\cos(k_y)$ as already outlined by Mazin \textit{et al.} Ref.~\onlinecite{Mazin:2010a}.

\section{Subtraction of the continuum}
\label{Asec:continuum}
The fluctuation response is superposed on the particle-hole continuum that essentially reflects symmetry-resolved transport properties \cite{Devereaux:2007}. Since the contribution of the fluctuations is relatively strong here they can be isolated with little uncertainty. The simplest way is to use the continuum  at or slightly above the crossover temperature $T_f$ and subtract it from all spectra measured below $T_f$. This was sufficient for ErTe$_3$ \cite{Eiter:2013} but created negative intensities in the case of $\rm La_{2-x}Sr_xCuO_4$ \cite{Tassini:2005}. Here, we wish to compare the temperature dependence of the fluctuations to a theoretical prediction and have to improve on the subtraction of the continuum. To this end we make the analytical phenomenology for the $B_{1g}$ continuum temperature dependent in a way that yields $\Gamma_{0,B1g}(T)\propto \rho(T)$. This seems sensible since the proportionality holds for the $A_{1g}$ results in the entire temperature range above $T_{\rm SDW}$ and for the $B_{1g}$ spectra above $T_f$. Fig.~\ref{fig:synthetic_e-h} shows the steps and checks necessary for the procedure. The analytical function used reads 
\begin{eqnarray}
  \label{eq:continuum}
  \chi^{\prime\prime}_{\rm cont}(\Omega,T) &=& [\alpha_1+\alpha_2\cdot T] \tanh\left(\frac{\Omega}{\tilde\Gamma_0(T)}\right) + \nonumber\\
  & &[\beta_1+\beta_2\cdot T]\left(\frac{\Omega}{\tilde\Gamma_0(T)}\right)
\end{eqnarray}
which obeys $\chi^{\prime\prime}_{\rm cont}(-\Omega,T)=-\chi^{\prime\prime}_{\rm cont}(\Omega,T)$ as required by causality. $\alpha_1$, $\alpha_2$, $\beta_1$ and $\beta_2$ depend only on
doping $x$. For $x=0.025$ we used $\alpha_1=0.82379$, $\alpha_2=-0.00138$, $\beta_1=-0.00923$, and $\beta_2=0.00028$. $\tilde\Gamma_0(T)$ is a fitting parameter that is selected in a way that the inverse slope $\Gamma_c(0,T)$ of $\chi^{\prime\prime}_{\rm cont}(\Omega,T)$ follows the resistivity (orange diamonds in Fig.~\ref{fig:synthetic_e-h}\,d). If a constant continuum is used the fluctuations can be isolated in a
qualitatively similar fashion. However, the experimental data in Fig.~2 vary more
slowly close to $T_s$.

Below $T_s$ the uncertainties increase since surface layers accumulate rapidly in the presence of twin boundaries where the surface assumes a more polar character. This can be seen directly in Fig.~\ref{fig:T-spot}\,(c).

\section{Initial slope}
\label{Asec:slope}

\begin{figure}[tbp]
  \centering
  \includegraphics[width=1.0\columnwidth]{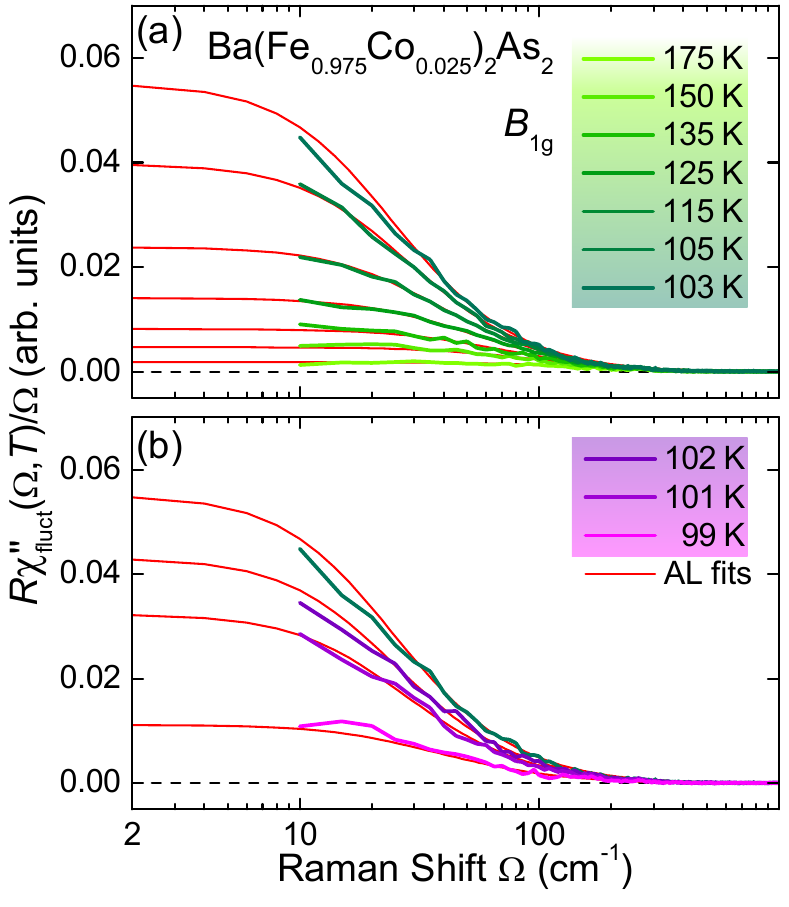}
  \caption{Initial slope of the Raman response of ${\rm Ba(Fe_{1-x}Co_{x})_2As_2}$ ($x=0.025$) below $T_f$. This figure is a reproduction of Fig.~\ref{fig:2.5AL} with the fluctuation response divided by the energy $\Omega$ plotted against a logarithmic energy scale.
  }
  \label{fig:R0/W}
\end{figure}

For being a causal function the Raman response is antisymmetric and, as long as there is no gap, linear around the origin. Then Eq.~\eqref{eq:slope} can be approximated as
\begin{eqnarray}
  \label{eq:linear_slope}
  \tau_{0,\mu}(T) &=& \left.\left(\frac{\partial
                 R\chi^{\prime\prime}_\mu(\Omega,T)}{\partial\Omega}\right)\right|_{\Omega=0} \nonumber\\
                  &=& \lim_{\Omega\rightarrow 0} \left(\frac{R\chi^{\prime\prime}_\mu(\Omega,T)}{\Omega}\right).
\end{eqnarray}
The temperature dependence (not the magnitude) of the initial slope can then directly be read off a graph if the response is divided by the energy $\Omega$ and plotted against a logarithmic energy scale.

If $R$ was known $\tau_0(T)$ could be determined directly. With $R$ unknown only the relative change can be derived in this way. Fig.~\ref{fig:R0/W} shows that the fits reproduce the overall data rather well at low energy. The phenomenological curves can be extended to arbitrarily low energies providing a simple way to directly visualize the temperature dependence of $\tau_0(T)$. Fig.~\ref{fig:R0/W} shows also that the experimental data close to zero energy are not very stable. This problem arises from accumulating surface layers and the influence of the laser line. Therefore, the error bars become excessively large if the slope is directly extracted from the data. Here we use a wide spectral range to improve the reproducibility.

\end{appendix}

%%%%%%%%%%%%%%%%%%%%%%%%%%%%%%%%%%%%%%%%%%%%%%%%%%%%%%%%%%%%%%%%%%%%%%%%%%%%%%%%%%%%%%%%%%%%%%%%%%%%%%
%\begin{center}\begin{center}\begin{flushleft}\end{flushleft}\end{center}\end{center}
%\subsection{A Subsection}
%\bibliography{D:/!papers/!bib/literatureR2}
%\bibliography{literatureR2}
%\bibliographystyle{prsty}
%
%%%%%%%%%%%%%%%%%%%%%%%%%%%%%%%%%%%%%%%%%%%%%%%%%%%%%%%%%%%%%%%%%%%%%%%%%%%%%%%%%%%%%%%%%%%%%%%%%%%%%%

\end{document}